\documentclass[10pt,letterpaper]{article}

\usepackage{opex3}
\usepackage{color}
\usepackage{cite}
\usepackage{amsmath,amsfonts,amssymb}

\newcommand{\be}{\begin{equation}}
\newcommand{\ee}{\end{equation}}

\def\W0{\omega_{0}}

\def\te{t_{e}}
\def\tl{t_{l}}

\def\in{\text{in}}
\def\out{\text{out}}

\begin{document}


\title{Topologically robust transport of entangled photons in a 2D photonic system}
\author{Sunil Mittal,$^{1,2,*}$ Venkata Vikram Orre,$^{1,2}$ and Mohammad Hafezi$^{1,2,3}$}
\address{$^{1}$Joint Quantum Institute, NIST/University of Maryland, College Park MD 20742, USA \\
         $^{2}$Department of Electrical and Computer Engineering and IREAP, \\ University of Maryland, College Park MD 20742, USA \\
         $^{3}$Kavli Institute of Theoretical Physics, Santa Barbara, CA 93106, USA}
\email{$^*$mittals@umd.edu}

\begin{abstract}
We theoretically study the transport of time-bin entangled photon pairs in a two-dimensional topological photonic system of coupled ring resonators. This system implements the integer quantum Hall model using a synthetic gauge field and exhibits topologically robust edge states. We show that the transport through edge states preserves temporal correlations of entangled photons whereas bulk transport does not preserve these correlations and can lead to significant unwanted temporal bunching or anti-bunching of photons. We study the effect of disorder on the quantum transport properties; while the edge transport remains robust, bulk transport is very susceptible, and in the limit of strong disorder, bulk states become localized. We show that this localization is manifested as an enhanced bunching/anti-bunching of photons. This topologically robust transport of correlations through edge states could enable robust on-chip quantum communication channels and delay lines for information encoded in temporal correlations of photons.
\end{abstract}

\ocis{(270.0270) Quantum optics; (230.4555) Coupled resonators.}


\section{Introduction}
Recently, there has been a significant surge of interest in investigating topological photonic systems implemented using synthetic gauge fields \cite{Hafezi2014,Lu2014}. These systems exhibit topologically robust edge states which are protected against backscattering caused by disorder. Topological edge states have been observed in photonic systems,  using coupled ring resonators \cite{Hafezi2011, Hafezi2013} and helical waveguides \cite{Rechtsman2013} in the optical domain, and  also using metamaterials \cite{Wang2009,Khanikaev2013,Ma2015,Cheng2016,Chen2014}  in the microwave domain. Many other interesting platforms have been proposed to realize photonic edge states \cite{Fang2012,Umucalilar2011,Kraus2012,Tzuang2014,Ningyuan2015}.  These states are characterized by a topologically invariant integer, the winding number, which has been measured recently in photonic systems \cite{Wenchao2015,Zeuner2015,Hafezi2014B,Mittal2016}. Moreover, the topological robustness of photonic edge states against disorder has been quantitatively established,  which merits their use as robust on-chip communication channels and delay lines \cite{Wang2009,Mittal2014}. However, investigations of topological robustness in photonic systems have so far relied on observing classical transport parameters such as transmission and delay statistics. Here, we study quantum transport properties of non-classical light in a two-dimensional (2D) topological photonic system. In contrast to previous works where the input was a classical state, here, the input is a quantum state of two time-bin entangled photons \cite{Brendel1999,Marcikic2002, Jayakumar2014}.

Our topological system is a 2D lattice of coupled ring resonators, shown in Fig.~\ref{fig:1}(a). A uniform synthetic magnetic field is introduced in this system by appropriately positioning the ring resonators [Fig.~\ref{fig:1}(b)] \cite{Hafezi2011,Hafezi2013}. This system simulates the integer quantum Hall model and its transmission spectrum is divided into bulk bands separated by topologically non-trivial edge bands where the corresponding states propagate around the edge of the system \cite{Hafezi2011,Hafezi2013}.  We show that the transport through edge band preserves  temporal correlations of the input photons whereas bulk transport significantly distorts these correlations, even in the absence of disorder.  In particular, for bulk transport, temporally anti-bunched photons at the input can bunch at the output and vice-versa. We compare these results with the case of separable two-photon states and show that this bunching/antibunching is more prominent for the entangled two-photon states, indicating that the correlated quantum states are more fragile. Furthermore, we study the effect of disorder on the quantum transport. Using the $\Psi^{+}$ Bell state as an example, where the two photons arrive at the input at different times, we show that disorder leads to temporal bunching of photons at the output and, this bunching increases with disorder strength.

We begin with a brief description of our system of coupled ring resonators and the two-photon input state. Subsequently, we analyze edge and bulk transport of maximally entangled Bell states in a system without disorder and contrast this to transport of separable states. Finally, we study the effect of disorder on the quantum transport.

\section{Entangled photons in a quantum Hall model}

\begin{figure}[h]
 \centering
 \includegraphics[width=0.98\textwidth]{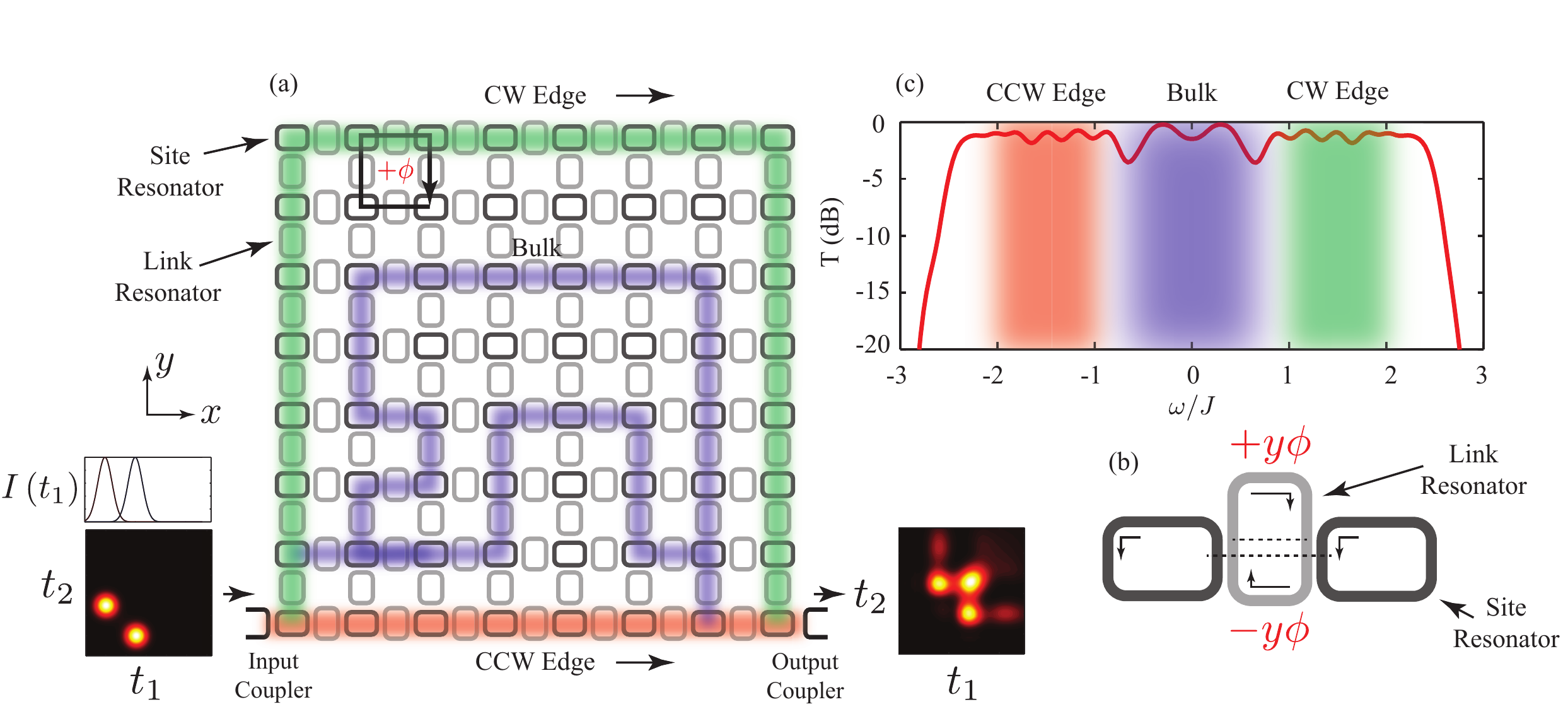}
 \caption{(a) Schematic of a 2D lattice of coupled ring resonators implementing the integer quantum-Hall model. Site resonators (black) are coupled using link resonators (grey). The lattice is coupled to input and output waveguides. Edge states transport is confined along the lattice boundary whereas the bulk states follow different paths through the bulk of the lattice. A time-bin entangled photon pair is coupled to the lattice at  input and the output temporal correlations are examined. An example single photon temporal wavefunction and the two-photon correlation function is shown at the input and the output.  (b) A vertical shift of link resonator introduces direction dependent hopping phase and hence synthetic magnetic field for photons. Photons hopping along right experience a longer path and hence an extra phase compared to photons hopping along left. (c) Single-photon transmission spectrum (solid red line) for a pure 8$\times$8 lattice. CW, CCW Edge and bulk bands are shaded in green, red and blue, respectively. In this paper, we use the input/output coupling rate to be same as the coupling rate $J$ between site resonators. }
 \label{fig:1}
\end{figure}

The 2D lattice of evanescently coupled ring resonators is shown in Fig.~\ref{fig:1}(a) \cite{Hafezi2011,Hafezi2013}. The site resonators (shown in black) are coupled using link resonators (in grey).  To introduce a synthetic magnetic field, the link resonators connecting site resonators are vertically shifted [Fig.~\ref{fig:1}(b)]. This vertical shift leads to a direction dependent hopping phase, i.e., a photon hopping to the right travels a longer path in the link resonator and hence accumulates an extra phase $\phi$, compared to a photon hopping to the left. This lattice implements the quantum spin-Hall model, where the two pseudo-spins correspond to clockwise and counterclockwise circulation of photons in the site resonators. However, by appropriately choosing the input port [Fig.~\ref{fig:1}(a)], we can selectively excite a single pseudo-spin component and simulate the integer quantum Hall model with the tight-binding Hamiltonian given as
\begin{equation}
 H =  \sum\limits_{x,y} \omega_{0} ~\hat{a}^{\dag}_{x,y} \hat{a}^{}_{x,y} - J \left(\hat{a}^{\dag}_{x+1,y} \hat{a}^{}_{x,y} e^{iy\phi} + \hat{a}^{\dag}_{x,y} \hat{a}^{}_{x+1,y} e^{-iy\phi} + \hat{a}^{\dag}_{x,y+1} \hat{a}^{}_{x,y} + \hat{a}^{\dag}_{x,y} \hat{a}^{}_{x,y+1} \right),
\end{equation}
where $\omega_{0}$ is the ring resonance frequency, $J$ is the coupling rate between neighboring lattice sites and $\phi$ is the synthetic magnetic flux threading a single plaquette.  $\hat{a}^{\dag}_{x,y}$ and $\hat{a}^{}_{x,y}$ are the photon creation and annihilation operators, respectively, at the lattice site $\left(x,y\right)$. We have specifically chosen the Landau gauge where the magnetic phase is associated only with hopping along $x$-direction and it is a linear function of the row index $y$. For simplicity, we choose $\omega_{0} = 0$. Moreover, to elucidate the topological protection of edge states against disorder, we neglect the effect of loss in the resonators which can lead to decoherence of the entangled state, in addition to disorder. Also, in experimental realization of this system, the effect loss is very small compared to that of disorder \cite{Hafezi2013, Mittal2014}.

Figure~\ref{fig:1}(c) shows the simulated single-photon transmission spectrum for a 8$\times$8 lattice, with a magnetic flux $\phi = \frac{2\pi}{4}$ per plaquette. The transmission spectrum is divided into bulk bands separated by edge bands \cite{Hatsugai1993B}. The edge bands (shaded in green and red) are associated with topologically non-trivial edge states circulating clockwise (CW) and counterclockwise (CCW) along the system boundary. On the other hand, states in the bulk band (shaded in blue)  occupy the bulk of the lattice \cite{Hafezi2011,Hafezi2013}.

\begin{figure}[h]
 \centering
 \includegraphics[width=0.98\textwidth]{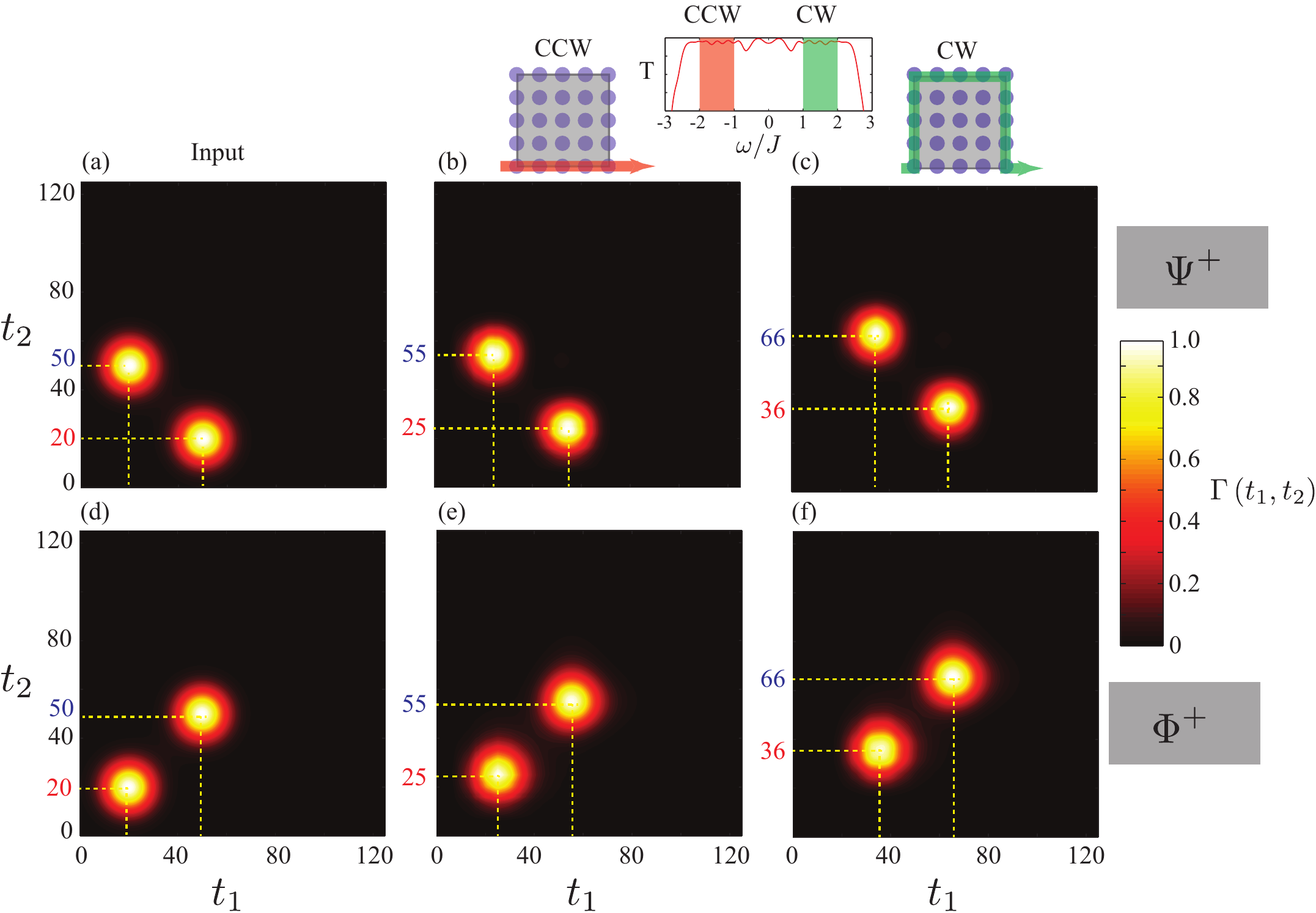}
 \caption{(a) Time-correlation $\Gamma(t_{1},t_{2})$ for $\Psi^{+}$ input state, with $\sigma = 10~T_{0}$ and delay $\tau = 30~T_{0}$, where $T_{0} = 1/J$. (b,c) Simulated correlation function at the output port of a 8$\times$8 lattice for CCW and CW edge states, respectively. The delay incurred in the edge states shifts the correlation function diagonally but correlation of the input state is preserved. The centres of the two time-bins are marked with dashed yellow lines. (d-f) Results for the input state $\Phi^{+}$. Insets show the transmission spectrum and the path followed by edge states. $\Gamma\left(t_{1},t_{2} \right)$ is normalized such that the maximum is unity.}
 \label{fig:2}
\end{figure}

At the input of this lattice, we couple a time-bin entangled two-photon state of the form
\be
\left|\psi\right> = \int^{\infty}_{-\infty} \int^{\infty}_{-\infty} dt_{1} dt_{2} \psi(t_{1},t_{2};t_{e},t_{l}) \hat{a}^{\dag}(t_{1}) \hat{a}^{\dag}(t_{2}) \left|0\right>,
\ee
where $(t_{e})$  and  $(t_{l})$ correspond to the early and late time bins in which the photons could arrive and $\hat{a}^{\dag}(t)$ is the photon creation operator at time $t$. $\psi(t_{1},t_{2};t_{e},t_{l})$ is the two-photon temporal wavefunction and is symmetric under exchange of photons. Note that both the photons are centered around the same carrier frequency and have same polarization, in the plane of ring resonators. Here, we consider the maximally entangled states - the Bell states. For example, the $\Psi^{+}$ state is written as
\be
\left| \Psi^{+}\right> = \frac{1}{\sqrt{2}} \left( \left|e\right>_{1} \left|l\right>_{2}  +   \left|l\right>_{1} \left|e\right>_{2}  \right),
\ee
where $\left|e\right>_{1,2}$ and $\left|l\right>_{1,2}$ represent the single-photon states in early and late time bins, respectively. It corresponds to a situation when one photon arrives in the early time-bin $(\te)$ and the other in the late bin $(\tl)$. The early/late time bins can be considered as "0/1" logic values of a qubit. Similarly, the other two Bell states symmetric under exchange of photons are
\begin{align}
\left| \Phi^{+} \right> &= \frac{1}{\sqrt{2}} \left( \left|e \right>_{1} \left|e \right>_{2}  +  \left|l \right>_{1} \left|l \right>_{2}  \right) \\
\left| \Phi^{-} \right> &= \frac{1}{\sqrt{2}} \left( \left|e \right>_{1} \left|e \right>_{2}  -  \left|l \right>_{1} \left|l \right>_{2}  \right).
\end{align}
These are the symmetric and antisymmetric combinations of the two scenarios when both the photons arrive early or both arrive late.  The fourth Bell state  $\Psi^{-}$ is not considered here because it is antisymmetric under exchange of photons. These time-bin entangled two-photon states can be realized in various systems, for example, using spontaneous parametric down conversion or quantum dots \cite{Brendel1999,Marcikic2002,Jayakumar2014}.

Assuming the input single-photon temporal wavefunctions are Gaussian, the two-photon wavefunction for $\Psi^{+}$ state is given by
\be
\Psi^{+}(t_{1},t_{2};t_{e},t_{l}) =  \mathcal{A} \left[ \text{exp} \left(-\frac{\left(t_{1}-t_{e}\right)^2}{2\sigma^{2}}\right) \text{exp} \left(-\frac{\left(t_{2}-t_{l}\right)^2}{2\sigma^{2}}\right) + \text{exp} \left(-\frac{\left(t_{1}-t_{l}\right)^2}{2\sigma^{2}}\right) \text{exp} \left(-\frac{\left(t_{2}-t_{e}\right)^2}{2\sigma^{2}}\right) \right],
\ee
where $\sigma$ characterizes the single-photon temporal pulsewidth and $\mathcal{A}$ is the normalization factor. Similarly, we can write the wavefunctions for the $\Phi^{\pm}$ states. The temporal two-photon correlation function $\Gamma \left(t_{1},t_{2}\right) = \left|\psi \left(t_{1},t_{2}\right) \right|^{2}$ is defined as the probability of finding one photon at time $t_{1}$ and the other photon at time $t_{2}$. The correlation functions for the $\Psi^{+}$ and $\Phi^{+}$ states at the input are shown in Figs.~\ref{fig:2}(a), \ref{fig:2}(d). Throughout this paper, we use $\sigma = 10 T_{0}$, where the inverse of the coupling rate $T_{0} = 1/J$ is the relevant time scale in the system. This choice of temporal pulsewidth ensures that the single-photon spectral width $(1/\sigma)$ is much less than the edge/bulk bandwidths and therefore, allows us to study the physics of each band separately. We also select the delay between the two time bins $\tau = \tl - \te = 30 T_{0}$, such that they have negligible overlap at the input. The maximum value of $\tau$ is set by the system size, i.e., $\tau < N_{x} N_{y} T_{0}$, where $N_{x,y}$ is the number of site resonators in the $x,y$ direction. In our case, $N_{x} = N_{y} = 8$.

\section{Results on transport properties of entangled photons}

Now, we analyze transport of these maximally entangled two photons in a system without disorder. For a linear system with no interactions, the two-photon temporal wavefunction can be calculated using the single-photon temporal wavefunctions at the output (see Appendix). Figures \ref{fig:2}(a)-\ref{fig:2}(c) show the simulated temporal correlation function at the input and the output for $\Psi^{+}$ state as the input, with central carrier frequency in the CCW and CW edge bands (at $\omega = \pm 1.5J$).  We see that for both the CCW and CW edge states, the correlation function is translated along the diagonal by the delay incurred in these paths which is equal to $5 T_{0}$ and $16T_{0}$, respectively. At the same time, the temporal correlation of input photons is clearly preserved for transport along the edge states. Similarly, Figs.~\ref{fig:2}(d)-\ref{fig:2}(f) show results for the $\Phi^{+}$ state where the correlations are also maintained at the output. We find similar results for $\Phi^{-}$ state. While Fig.~\ref{fig:2} presents results for a single input frequency, the behavior remains same across the band. Furthermore, the correlations are preserved irrespective of system size and also the extent of single-photon wavefunction overlap at the input.

\begin{figure}[h]
 \centering
 \includegraphics[width=0.98\textwidth]{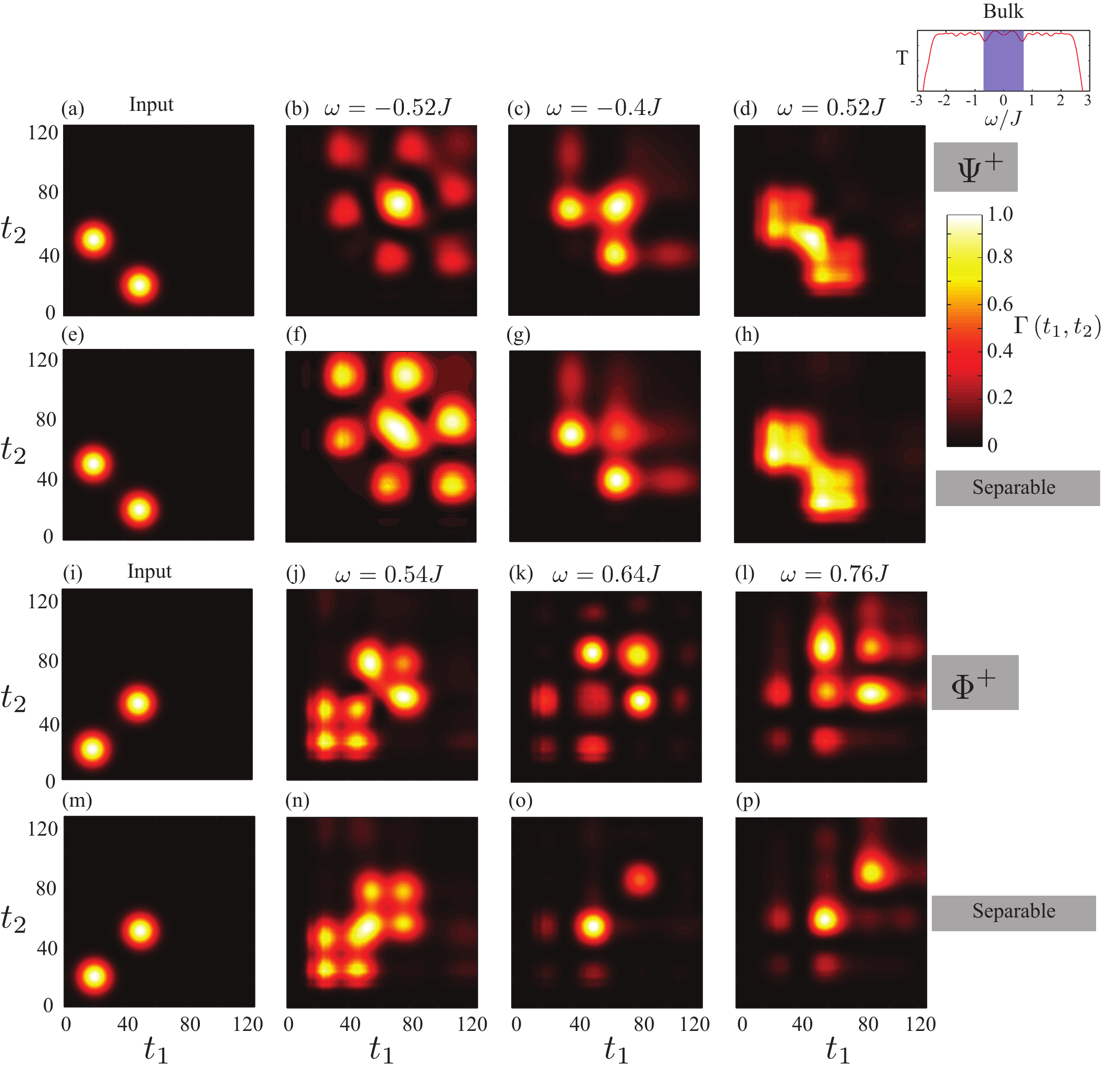}
 \caption{(a-d) Time-correlation function at the input and the output of the lattice for $\Psi^{+}$ state and three different input frequencies in the bulk band, $\omega = (-0.52, -0.4, 0.52) J$. The profile is dictated largely by the input excitation frequency and the two photons can bunch at the output even when they are well separated at the input. (e-h) Correlation for the separable state corresponding to the input frequencies in (a-d). For the separable state, the bunching is much less than that for the entangled state. (i-p) Simulation results for $\Phi^{+}$ and the corresponding separable state, where the photons are bunched at the input and can anti-bunch at the output after propagating through bulk states. These results show that the quantum state of two entangled photons is more fragile than the separable state.}
 \label{fig:3}
\end{figure}

To show that this preservation of temporal correlations in a 2D system is not trivial, we analyze transport through the bulk band. Unlike edge transport, the output correlation function for bulk transport varies with the input excitation frequency in the band and can also be significantly distinct from the input. Figures~\ref{fig:3}(a)-\ref{fig:3}(d) show the simulated output for $\Psi^{+}$ input state, with $\omega = \left(-0.52, -0.4,~0.52\right) J $. Interestingly, we find that the two photons can bunch at the output even though they were anti-bunched at the input. We further contrast the transport of entangled photons to that of a separable two-photon state with distinguishable photons, i.e., when there are no interference effects and the transport is essentially single-particle physics. For the separable state corresponding to state $\Psi^{+}$, the two-photon wavefunction is $\psi\left(t_{1},t_{2};t_{e},t_{l}\right) = \phi_{1}\left(t_{1}-t_{e}\right) \phi_{2}\left(t_{2}-t_{l}\right)$, where $\phi_{i}\left(t_{i}\right)$ is the single-photon temporal wavefunction. For comparison, we  symmetrize the two-photon correlation function for separable state as $\Gamma\left(t_{1},t_{2}\right) = \frac{1}{2} \left( \left|\phi_{1} \left(t_{1}-t_{e}\right) \phi_{2} \left(t_{2}-t_{l}\right)\right|^{2} + \left|\phi_{1} \left(t_{1}-t_{l}\right) \phi_{2} \left(t_{2}-t_{e}\right)\right|^{2} \right)$. The simulation results are shown in Figs.~\ref{fig:3}(e)-\ref{fig:3}(h) and we see that for the separable state, bunching is much less pronounced than that of the entangled state. Similarly, Figs.~\ref{fig:3}(i)-\ref{fig:3}(l) show simulation results for $\Phi^{+}$ state where the two photons are bunched at the input, but are anti-bunched at the output. The results for a separable state corresponding to $\Phi^{+}$ state are shown in Figs.~\ref{fig:3}(m)-\ref{fig:3}(p) and in this case we find that the anti-bunching is less prominent than the entangled state. This clearly demonstrates that the correlated quantum state of two photons is much more fragile than the separable state.

The temporal bunching/anti-bunching of photons seen here can be compared to spatial bunching/anti-bunching of photons which has been observed in the case of quantum walks of spatially correlated photons \cite{Bromberg2009,Lahini2010,Peruzzo2010, Crespi2013,Giuseppe2013,Poulios2014}. Specifically, quantum walks of photons have been implemented using arrays of beam-splitters or continuously coupled waveguides and, these systems have multiple input and output ports \cite{Sansoni2012,Schreiber2012,Crespi2013}. The correlated photons are coupled to different input ports and quantum walk in the system leads to spatial bunching/anti-bunching of photons at the output, depending on the choice of input excitation ports and relative phase between them. In contrast, our system has a single input and a single output port. But, each coupling region between the resonators is a beam-splitter and therefore, the transport of photons from input to the output by hopping this array of beam-splitters can be considered as a 2D spatial quantum walk of two photons. These spatial correlations of the two-photon quantum walk in the lattice manifest as temporal correlations at the output port.

\section{The effect of disorder}

\begin{figure}[h]
 \centering
 \includegraphics[width=0.98\textwidth]{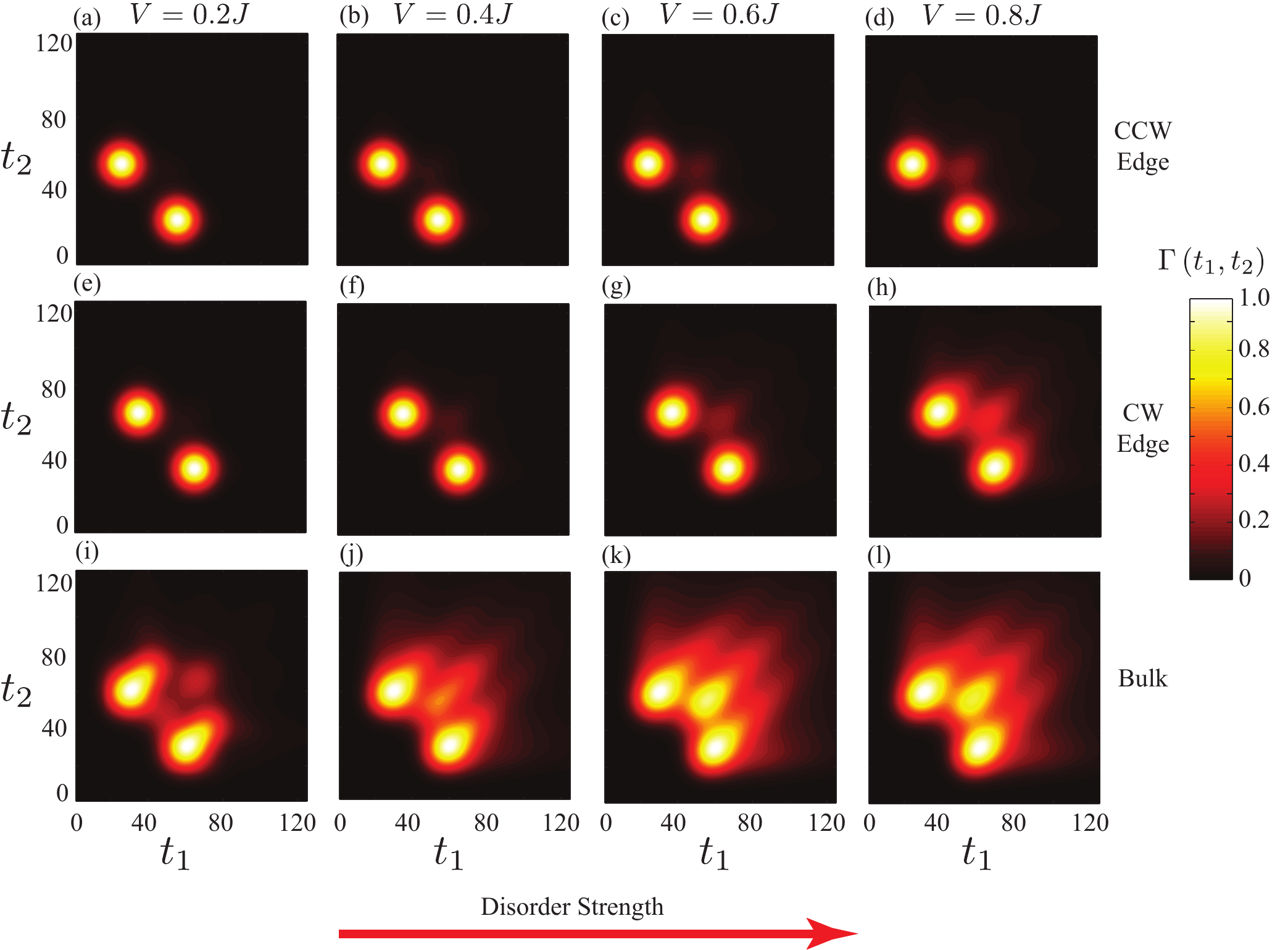}
 \caption{Correlation function at the output of a disordered system, for $\Psi^{+}$ input excitation in the (a-d) CCW edge band, (e-h) CW edge band and (i-l) for bulk band; for four different disorder strengths. As disorder increases, the bunching increases and is more significant for bulk band than the edge bands.}
 \label{fig:4}
\end{figure}

On-chip quantum information processing using time-bin entangled photons would require delay lines to temporally manipulate photons, for example, synchronize/desynchronize various sources and detectors of entangled photons. Although 1D coupled-resonator single-photon delay lines have been demonstrated \cite{Takesue2013}, they are plagued by disorder. Disorder leads to localization where transmission through the system falls exponentially with the length of the system \cite{Shayan2008, Mittal2014}. Disorder also limits time-bandwidth product in 1D systems because an increase in the delay, by increasing the number of resonators, comes at the cost of operational bandwidth \cite{Cooper2010}.

Using transmission and delay statistics of classical light, it has been shown that the edge states can be used to implement topologically robust delay lines which are protected against disorder induced localization \cite{Wang2009, Mittal2014}.  Here, we use non-classical, two-photon entangled states as the input and show that quantum transport through edge states is also topologically protected against localization whereas the bulk states localize. In particular, we consider a disorder in the form of on-site potential $V$ which is a result of difference in the site ring resonance frequencies $(\Delta \omega_{0})$ and is the most significant disorder which affects 1D coupled-resonator delay lines \cite{Shayan2008, Mittal2014}. We consider $\Psi^{+}$ input state as an example, where the two photons are anti-bunched at the input, and show that localization manifests as temporal bunching of photons at the output.

Figures~\ref{fig:4}(a)-\ref{fig:4}(h) shows simulation results at the output with input excitation in the CCW/CW edge and bulk bands, for different disorder strengths $V = \left(0.2,~0.4,~0.6,~0.8\right) J$. The correlation functions have been averaged over the corresponding bands as well as over 500 realizations of random disorder in the lattice. We can clearly see that the edge states maintain the correlation function of the input.  We see some bunching only at very strong disorder strengths of $U > 0.6 J$.  Also, note that the bunching is slightly more prominent for CW edge band because it travels a longer path through the lattice and is therefore more disposed to disorder. Ideally, we expect that the edge states would show no bunching at all. However, when the disorder strength is comparable to the coupling rate $J$, as is the case here, the edge bandwidth shrinks and the edge states start to loose their topological protection. On the contrary, for bulk band [Figs.~\ref{fig:4}(i)-\ref{fig:4}(l)], even small disorder leads to significant bunching and the bunching probability increases as the disorder increases.

\begin{figure}
 \centering
 \includegraphics[width=0.98\textwidth]{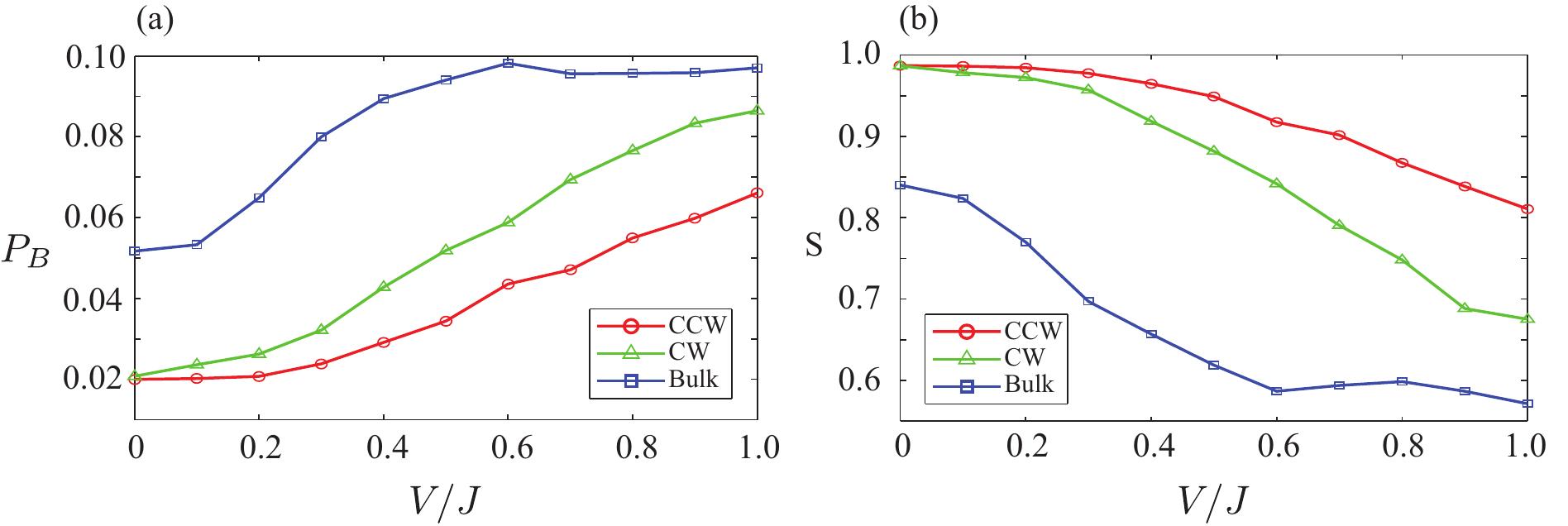}
 \caption{(a) Calculated probability of bunching for $\Psi^{+}$ input state, with excitation in the edge and the bulk bands, as a function of disorder. Bunching is more prominent for bulk states and increases with disorder strength. (b) Similarity between the input and the output correlation. Even in the presence of strong disorder, the output correlation for edge states is very similar to that of the input. }
 \label{fig:5}
\end{figure}

To quantify the effect of disorder on quantum transport, we calculate the normalized probability of bunching at the output for $\Psi^{+}$ Bell state which is anti-bunched at the input, defined as
\begin{equation}
  P_{B} = \frac{\int^{\infty}_{-\infty} \int^{\infty}_{-\infty} dt_{1} dt_{2} \delta(t_{2}-t_{1}\pm \epsilon) \Gamma(t_{1},t_{2})}{\int^{\infty}_{-\infty} \int^{\infty}_{-\infty} dt_{1} dt_{2} \Gamma(t_{1},t_{2})},
\end{equation}
where the two photons arriving in a time window of $\epsilon = \sigma/2$ are considered to be bunched. Figure~\ref{fig:5}(a) shows $P_{B}$ as a function of disorder strength. The probability of bunching for the edge states is much less compared to that of the bulk states. Also, the probability of bunching increases with disorder strength.  We further look at the similarity of the output correlation function to the input correlation, defined as
\begin{equation}
S = \frac{\left( \int^{\infty}_{-\infty} \int^{\infty}_{-\infty} dt_{1} dt_{2} \sqrt{\Gamma_{\text{out}}(t_{1},t_{2}) \Gamma_{\text{in}}(t_{1},t_{2})} \right)^{2}} {\int^{\infty}_{-\infty} \int^{\infty}_{-\infty} dt_{1} dt_{2} \Gamma_{\text{out}}(t_{1},t_{2})  \int^{\infty}_{-\infty} \int^{\infty}_{-\infty} dt_{1} dt_{2} \Gamma_{\text{in}}(t_{1},t_{2})},
\end{equation}
where $\Gamma_{\text{in,out}}$ are the correlation functions at the input and the output. Figure~\ref{fig:5}(b) shows the similarity $S$ as a function of disorder strength. As expected, the correlation function at the output for edge transport resembles the input more as compared to that for bulk transport.

We have also simulated transport of $\Phi^{\pm}$ states through disordered system. Similar to $\Psi^{+}$ state, we find that the edge bands preserve temporal correlations. On the other hand, for bulk transport, the two-photon anti-bunching increases with disorder, unlike $\Psi^{+}$ state where disorder leads to bunching.  Note that for $\Phi^{\pm}$ states, the photon are bunched at the input. Also, as we saw for $\Psi^{+}$ state, the similarity between the input and the output correlation decreases with increasing disorder strength.

\section{Summary}
To summarize, we have studied transport of time-bin entangled photon pairs in a 2D topological photonic system. We have shown that the quantum transport through edge band preserves temporal correlations of time-bin entangled photons. In contrast, transport in the bulk band significantly distorts these correlations, and can result in spurious bunching/ anti-bunching of photons. Using $\Psi^{+}$ state as an example, we have shown that disorder in the system leads to increased bunching for bulk states whereas for edge states, this bunching is significantly less. This topological robustness of edge states in preserving quantum correlations of entangled photons could be used to realize robust on-chip communication channels and delay lines for the entangled photons. Similar physics could be observed in the microwave domain \cite{Anderson2016} and also in exciton-polariton systems \cite{Sala2015}. Moreover, it would be interesting to investigate the transport of entangled photons in synthetic dimensions \cite{Yuan2016, Ozawa2016}.


\begin{center} \textbf{Appendix} \end{center}

In this section we show that for the linear 2D system of coupled ring resonators considered here, the two-photon temporal correlation function at the output can be calculated using the single-photon temporal wavefunction at the output. We begin by writing the general form of the time-bin entangled state of two indistinguishable photons as
\be
\left|\psi_{\in}(t_{e},t_{l}) \right> = \mathcal{A} \int^{\infty}_{-\infty} \int^{\infty}_{-\infty} dt_{1} dt_{2} \psi_{\in}(t_{1},t_{2};t_{e},t_{l}) \hat{a}^{\dag}(t_{1}) \hat{a}^{\dag}(t_{2}) \left|0\right>.
\ee
Here $t_{e,l}$  correspond to the early and late time bins and $\mathcal{A}$ is a normalization factor. $\psi_{\in}(t_{1},t_{2};t_{e},t_{l})$ is the two-photon temporal wavefunction which is symmetric under exchange of photons and is given by
\be
\psi_{\in}(t_{1},t_{2};t_{e},t_{l}) = \frac{1}{\sqrt{2}} \left(\phi_{\in,1}(t_{1}-t_{e}) \phi_{\in,2}(t_{2}-t_{l}) + \phi_{\in,1}(t_{1}-t_{l}) \phi_{\in,2}(t_{2}-t_{e}) \right),
\ee
where $\phi_{\in,i}(t_{i}-t_{e(l)})$ is the single-photon temporal wavefunctions corresponding to the photon arriving in the early(late) time bin. Using 2D Fourier transform, the two-photon temporal wavefunction can be rewritten in frequency domain as
\be
\psi_{\in}\left(t_{1},t_{2};t_{e},t_{l}\right) = \frac{1}{2\pi} \int^{\infty}_{-\infty} \int^{\infty}_{-\infty} d\omega_{1} d\omega_{2} \tilde{\psi}_{\in}(\omega_{1},\omega_{2}) e^{-i\omega_{1}t_{1}} e^{-i\omega_{2}t_{2}},
\ee
and $\tilde{\psi}_{\in}(\omega_{1},\omega_{2})$ is now the two-photon spectral wavefunction at the input. Using (10),
\be
\tilde{\psi}_{\in}(\omega_{1},\omega_{2}) = \tilde{\phi}_{\in,1}(\omega_{1}) \tilde{\phi}_{\in,2}(\omega_{2}) \left[ \text{exp}  \left(i \omega_{1} t_{e} + i \omega_{2} t_{l} \right) + \text{exp}  \left(i \omega_{1} t_{l} + i \omega_{2} t_{e} \right) \right],
\ee
where $\tilde{\phi}_{\in,i}(\omega)$ is the Fourier transform of single particle temporal wavefunction $\phi_{\in,i} (t)$. Note that the two photons are centered around the same central frequency. For a Hamiltonian which does not have any nonlinear terms (e.g. of the form $\hat{a}^{\dag 2}\hat{a}^{2}$), there is no spectral mixing and the two-photon spectral function at the output is
\begin{equation}
\begin{split}
\tilde{\psi}_{\out} \left(\omega_{1},\omega_{2} \right)  &= S\left(\omega_{1}\right) S\left(\omega_{2}\right) \tilde{\psi}_{\in}(\omega_{1},\omega_{2}) \\
&=  \tilde{\phi}_{\out,1}(\omega_{1}) \tilde{\phi}_{\out,2}(\omega_{2}) \left[ \text{exp}  \left(i \omega_{1} t_{e} + i \omega_{2} t_{l} \right) + \text{exp}  \left(i \omega_{1} t_{l} + i \omega_{2} t_{e} \right) \right] ,
\end{split}
\end{equation}
where $S(\omega)$ is the transfer function of the system for a single photon at input frequency $\omega$, i.e., $\tilde{\phi}_{\out,i} \left( \omega \right) = S(\omega) \tilde{\phi}_{\in,i} \left( \omega \right)$ and can be calculated using the input-output formalism \cite{Hafezi2011}. Using the inverse Fourier transform, the two-photon temporal wavefunction at the output can be expressed as
\begin{equation}
\begin{split}
\psi_{\out} \left(t_{1},t_{2}; t_{e},t_{l} \right) &= \frac{1}{2\pi} \int^{\infty}_{-\infty} \int^{\infty}_{-\infty} d\omega_{1} d\omega_{2} \tilde{\psi}_{\out}(\omega_{1},\omega_{2}) e^{-i\omega_{1}t_{1}} e^{-i\omega_{2}t_{2}} \\
                                                  &= \frac{1}{\sqrt{2}} \left(\phi_{\out,1}(t_{1}-t_{e}) \phi_{\out,2}(t_{2}-t_{l}) + \phi_{\out,1}(t_{1}-t_{l}) \phi_{\out,2}(t_{2}-t_{e}) \right).
\end{split}
\end{equation}
Here $\phi_{\out,i}\left(t\right)$ is the inverse Fourier transform of $\tilde{\phi}_{\out,i} \left(\omega \right)$. Thus, we have proved that for a non-interacting Hamiltonian, the two-photon temporal wavefunction at the output can be computed from the single-photon wavefunction. Similar approach has been used earlier to study quantum walks of spatially correlated photons. \\


\noindent \textbf{Note} \\
\noindent During the preparation of this manuscript, we became aware of a similar work by Rechstman et. al. \cite{Rechtsman2016,Rechtsman2015}. In that work, a photonic Floquet topological system of coupled helical waveguides was studied and edge states were shown to preserve transport of path-entangled photons. \\

\noindent \textbf{Acknowledgments}  \\
\noindent We thank Alejandro Lobos and Michael Gullans for fruitful discussions. This work was supported by  AFOSR, ONR-YIP, ARO-MURI, NSF-PFC  at the JQI, and the Sloan Foundation. KITP is supported by NSF PHY11-25915.

\end{document}